# 2<sup>nd</sup> Response to "Feasibility of 3D reconstruction from a single 2D diffraction measurement"

Jianwei Miao and Chien-Chun Chen

Department of Physics and Astronomy and California NanoSystems Institute,

University of California, Los Angeles, CA 90095, USA.

Email: miao@physics.ucla.edu.

## ABSTRACT

We present our 2<sup>nd</sup> response to Thibault's commentary article [1] and his reply [2], in which he commented upon our ankylography paper [3] and our 1<sup>st</sup> response [4]. In this article, we further explain why we think Thibault's theoretical analysis is flawed and his interpretation of our experiment is incorrect. Furthermore, we provide a quantitative analysis and a numerical experiment to illustrate why ankylography can in principle be applicable to general samples. Finally, we present detailed procedures for our numerical experiment on ankylographic reconstructions, which uses the traditional HIO algorithm only with the positivity constraint [5]. We welcome anyone (including Thibault) interested in ankylography to perform numerical experiments and verify our results. We will be very happy to provide any help if needed.

# I. Flawed Theoretical Analysis in Refs. 1 and 2

First, the reason that we discussed oversampling *vs.* the autocorrelation function in ref. 4 is to illustrate the fact that *the reconstruction of an object is fundamentally different from the reconstruction of its autocorrelation function*. As a mater of fact, we can come up many counter-examples to show that, although the object can be retrieved from the oversampled diffraction intensities by using the phase retrieval algorithm, its autocorrelation function can *not* be directly obtained from the intensities [6]. In science, if a theory is in contradiction with many counter-examples, it is fair to call it flawed. Although Thibault acknowledged this limitation and his ignoring of the physical constraints in his theoretical analysis in ref. 2, the important and unfortunate fact is that his theory doesn't take them into account. Thus we think it is incorrect to draw general conclusions, such as "A second conclusion is that ankylography, as a general method, is doomed to failure most of the time." [page 9 in ref. 1] simply based on a flawed and overly-simplified theory, while completely ignoring our numerical simulation results present in ref. 3.

Second, Thibault stated in ref. 2 that "This freedom of choosing the sampling density is a direct consequence of Shannon's theorem: if Nyquist criterion is satisfied, the signal is completely determined by its samples. In your paper and in Figure 1 of your response, you chose a sampling substantially higher than the Nyquist criterion. Still, I want to emphasize that any sampling that is denser than Nyquist's criterion is redundant and contains no new information." Thibault may not realize that Shannon's theorem requires the signal and its Fourier-space function to be related by the Fourier transform and the inversion [7], which are linear and orthogonal. In ankylography, however, $\rho(\vec{r})$ and $F(\theta, \varphi)$ are related by Eq. (1) in ref. 3, which is neither the Fourier transform relationship nor orthogonal. In another words, even if having both the magnitudes and phases of $F(\theta, \varphi)$, one still can't directly obtain $\rho(\vec{r})$ from $F(\theta, \varphi)$, as the direct inversion of Eq. (1) doesn't exist. Furthermore, due to the missing of the phases, Eq. (1) in ref. 3 is not linear. This explains why Shannon's theorem is not directly applicable to ankylography.

Third, we noticed there are a few inaccurate statements in ref. 2. Specifically,

- Thibault stated that "Keeping this fact in mind, I hope that you now see how the Figure 1 in your response is completely equivalent to my Figure 1(e), the only difference being that I picked the coarsest sampling allowed by the Nyquist criterion". If carefully comparing Fig. 1(a) in ref. 4 with Fig. 1(e) in ref. 1, and reading the 2nd paragraph in page 5 in ref. 4, one would see the difference between the two figures. The difference lies in that the representation of grid points along the horizontal axis in Fig. 1e in ref. 1 is related to the super-resolution scheme, but not ankylography.

- Thibault stated that "To summarize: the maximum number of degrees of freedom that can be fixed by the data is the number of independent pixels in the autocorrelation." We can easily come up a counter-example to contradict this statement. Let's assume a 3D real object with the size of $a \times a \times a$ voxels. According to Thibault's argument [2], the number of independent voxels in the autocorrelation is $4a^3$. But according to ref. 8, we know that as long as the number of measured points $\geq 2a^3$ (*i.e.* $\sigma \geq 2$), the 3D object can be reconstructed from the diffraction pattern by using the iterative algorithm. From the reconstructed 3D object, one can then calculate the autocorrelation function, which indicates that these $4a^3$ points are not completely independent.

Finally, we will provide a quantitative analysis to explain why ankylography can in principle be applicable to general samples. We will also show numerical experiment results to support our conclusion. Let's assume that a coherent wave illuminates a 3D real object, $\rho(x, y, z)$. The far-field diffracted wave, $F(k_x, k_y, k_z)$, is oversampled on the Ewald sphere. We separate $F(k_x, k_y, k_z)$ into cosines and sines,

$$F(k_x, k_y, k_z) = A_{k_x, k_y, k_z} \exp(i\phi_{k_x, k_y, k_z})$$

$$= \sum_{x=-M}^{M} \sum_{y=-M}^{M} \sum_{z=-M}^{M} \rho(x,y,z) \exp\left[\frac{-2\pi i(k_x \cdot x + k_y \cdot y + k_z \cdot z)}{2N+1}\right]$$

$$\Rightarrow \begin{cases} A_{k_x, k_y, k_z} \cos(\phi_{k_x, k_y, k_z}) = \sum_{x=-M}^{M} \sum_{y=-M}^{M} \sum_{z=-M}^{M} \rho(x,y,z) \cos\left[\frac{2\pi(k_x \cdot x + k_y \cdot y + k_z \cdot z)}{2N+1}\right] \\ iA_{k_x, k_y, k_z} \sin(\phi_{k_x, k_y, k_z}) = -i \sum_{x=-M}^{M} \sum_{y=-M}^{M} \sum_{z=-M}^{M} \rho(x,y,z) \sin\left[\frac{2\pi(k_x \cdot x + k_y \cdot y + k_z \cdot z)}{2N+1}\right] \end{cases} \quad (1)$$

$$\forall k_x, k_y, k_z: \quad k_x^2 + k_y^2 + (k_z + N)^2 = N^2$$

where $(2M+1)^3$ is the size of the 3D object (*i.e.* support size), $(2N+1)^3$ is the size of the Fourier-space array in which the two hemi-sphere (*i.e.* Ewald sphere) are located, $A_{k_x, k_y, k_z}$ and $\phi_{k_x, k_y, k_z}$ are the magnitudes and phases of $F(k_x, k_y, k_z)$, and the diffraction angle is assumed to be 90°. Note that Eq. (1) is not the discrete Fourier transform as the reciprocal-space vectors on the Ewald sphere ($k_x$, $k_y$, $k_z$) are neither independent nor integers. In practice, we have to use the fast Fourier transform and its inversion for ankylographic reconstructions, which requires interpolating the measured data points onto a regular grid [3]. Here, we chose the Ewald sphere shell to be one pixel thick, which is a reasonable assumption as the thickness of the Ewald sphere shell is determined by the experimental parameters such as the energy resolution, the divergence and convergence angle of the incident beam. By only using the grid points within a Ewald sphere shell of 1 pixel thick, we write Eq. (1) into the matrix form,

$$BX = A$$

$$B = \begin{pmatrix} \cos\left[\dfrac{2\pi(k_{x1} \cdot x_{-M} + k_{y1} \cdot y_{-M} + k_{z1} \cdot z_{-M})}{2N+1}\right] & \cdots & \cos\left[\dfrac{2\pi(k_{x1} \cdot x_{M} + k_{y1} \cdot y_{M} + k_{z1} \cdot z_{M})}{2N+1}\right] \\ \vdots & & \\ \cos\left[\dfrac{2\pi(k_{xL} \cdot x_{-M} + k_{yL} \cdot y_{-M} + k_{zL} \cdot z_{-M})}{2N+1}\right] & \cdots & \cos\left[\dfrac{2\pi(k_{xL} \cdot x_{M} + k_{yL} \cdot y_{M} + k_{zL} \cdot z_{M})}{2N+1}\right] \\ 1 & \cdots & \cdots & 1 \\ -\sin\left[\dfrac{2\pi(k_{x1} \cdot x_{-M} + k_{y1} \cdot y_{-M} + k_{z1} \cdot z_{-M})}{2N+1}\right] & \cdots & -\sin\left[\dfrac{2\pi(k_{x1} \cdot x_{M} + k_{y1} \cdot y_{M} + k_{z1} \cdot z_{M})}{2N+1}\right] \\ \vdots & & \vdots \\ -\sin\left[\dfrac{2\pi(k_{xL} \cdot x_{-M} + k_{yL} \cdot y_{-M} + k_{zL} \cdot z_{-M})}{2N+1}\right] & \cdots & -\sin\left[\dfrac{2\pi(k_{xL} \cdot x_{M} + k_{yL} \cdot y_{M} + k_{zL} \cdot z_{M})}{2N+1}\right] \end{pmatrix} \quad (2)$$

$$X = \begin{pmatrix} \rho(x_{-M}, y_{-M}, z_{-M}) \\ \vdots \\ \rho(x_{M}, y_{M}, z_{M}) \end{pmatrix} \qquad A = \begin{pmatrix} A_1 \cos(\phi_1) \\ \vdots \\ A_L \cos(\phi_L) \\ A_0 \\ A_1 \sin(\phi_1) \\ \vdots \\ A_L \sin(\phi_L) \end{pmatrix}$$

$$\forall\, k_{xi}, k_{yj}, k_{zk} \quad i, j, k \in [-N, N]: \quad \left(N - \frac{1}{2}\right)^2 \le k_{xi}^2 + k_{yj}^2 + (k_{zk} + N)^2 < \left(N + \frac{1}{2}\right)^2$$

where *B, X* and *A* are $(2L+1) \times (2M+1)^3$, $(2M+1)^3 \times 1$ and $(2L+1) \times 1$ matrices, respectively, $(2L+1)$ is the number of non-centro-symmetrical grid points within a Ewald sphere shell, and the row of $(1 \ldots 1)$ in matrix *B* and $A_0$ in matrix *A* correspond to the centro-pixel. To facilitate our quantitative analysis, we generated two new matrices $B'$ and $X'$ by expanding *B* and padding zeros to *X*,

$$B' = \begin{pmatrix} \cos\left[\dfrac{2\pi(k_{x1}\cdot x_{-L} + k_{y1}\cdot y_{-L} + k_{z1}\cdot z_{-L})}{2N+1}\right] & \cdots & \cdots\cos\left[\dfrac{2\pi(k_{x1}\cdot x_{L} + k_{y1}\cdot y_{L} + k_{z1}\cdot z_{L})}{2N+1}\right] \\ \vdots & B & \vdots \\ -\sin\left[\dfrac{2\pi(k_{xL}\cdot x_{-L} + k_{yL}\cdot y_{-L} + k_{zL}\cdot z_{-L})}{2N+1}\right] & \cdots & \cdots-\sin\left[\dfrac{2\pi(k_{xL}\cdot x_{L} + k_{yL}\cdot y_{L} + k_{zL}\cdot z_{L})}{2N+1}\right] \end{pmatrix}$$

$$X' = \begin{pmatrix} 0 \\ \vdots \\ 0 \\ X \\ 0 \\ \vdots \\ 0 \end{pmatrix} \qquad \textit{such that} \qquad B'X' = A \tag{3}$$

where $B'$ and $X'$ become $(2L+1)\times(2L+1)$ and $(2L+1)\times1$ matrices, respectively. Mathematically, Eq. (3) is exactly equivalent to Eq. (2).

To answer the question of whether ankylography is applicable to general objects, we can examine the rank of square matrix $B'$. Let's calculate the rank of $B'$ by using a $7\times7\times7$ 3D object (*i.e.* $M = 3$). The Ewald sphere is embedded inside a $17\times17\times17$ array (*i.e.* $N = 8$). The number of non-centro-symmetrical grid points within a Ewald sphere shell of 1 pixel thick is 393 (*i.e.* $L = 392$) with $O_d = 1.14$. By using the standard Matlab codes, we calculated the rank of $B'$ to be 785 (*i.e.* matrix $B'$ has full rank). In this case, the number of unknown variables of the 3D object is 343 (*i.e.* $7^3$), and the number of unknown variables for the phases in Eq. (3) is 392. Therefore the total number of unknown variables is 735 which is smaller than the rank of $B'$, suggesting that the 3D object can in principle be obtained by solving Eq. (3). We also calculated the rank of $B'$ for a few different ankylography cases, and found that $B'$ always has full rank. Note that the calculation of the rank for matrix $B'$ will not be affected by noise, as noise is only present in matrix $A$ in Eq. (3). Our analysis above has clearly demonstrated that there is indeed a unique property about oversampling of diffraction intensities on the Ewald sphere, as it encodes information from all possible orientations of a 3D object (supplementary information in ref. 3) and hence enables us to reconstruct a 3D object from a spherical diffraction pattern alone.

To verify our quantitative analysis, we performed a numerical experiment on ankylographic reconstructions. For the purpose of independent confirmation, the numerical experiment was independently conducted by two members in our group, Russell Fung and Chien-Chun Chen who were not involved in the previous ankylography paper [3]. By using their own GHIO [9] and HIO codes, they both successfully reconstructed 3D objects from single 2D spherical diffraction patterns alone. The reason of using the HIO and GHIO algorithms in their simulations is to facilitate others who are interested in performing ankylographic reconstructions as the HIO algorithm can be easily implemented [5]. In the following we will present Chen's results while Fung's will be presented in a follow-up paper.

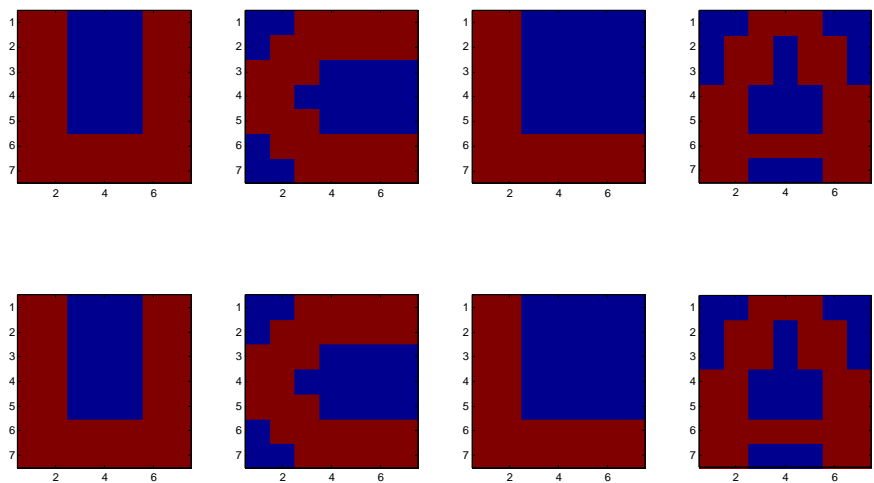

**Fig. 1** Ankylographic reconstruction of a 3D object from a simulated spherical diffraction pattern alone with $O_d = 1.14$. The upper panel shows the 1st, 3rd, 5th, 7th slices of the 3D reconstructed image. The lower panel shows the corresponding slices of the model, consisting four alphabet letters "U", "C", "L", and "A".

Fig. 1 (low panel) shows a 3D test object ($7^3$ voxels), which consists four alphabet letters "U", "C", "L", and "A". Each letter occupies one slice of the object and is separated from the neighboring letter by an empty slice. By padding zeros around the object and calculating the magnitudes of the Fourier transform, we generated four arrays with size of $101^3$, $23^3$, $19^3$ and $17^3$ voxels, respectively. We then kept the grid points within the Ewald sphere shell of 1 pixel thick and set the other grid points to zeros. The oversampling degree ($O_d$) of the four cases is 46.5, 2.18, 1.72 and 1.14, respectively (Tab. 1). Fig. 2 shows the distribution of the grid points within two hemi-spherical shells of 1 pixel thick. The size of the Fourier-space array is $17^3$ voxels, the number of non-centro-symmetrical grid points within a hemi-sphere shell is 393, and $O_d = 1.14$.

By using the HIO algorithm with only the positivity constraint [5], we reconstructed the 3D object for all four cases. Tab. 1 summarizes the reconstruction conditions where errorF and errorR are the Fourier-space and real-space R-factors, respectively, used for quantifying the reconstructions [9]. The smallest reconstruction errors for $O_d = 1.14$ is likely due to a larger number of iteration. We also noticed that the larger the oversampling degree ($O_d$), the higher the success rate of the reconstructions. Fig. 1 (upper panel) shows the reconstructed slices which are in excellent agreement with the original object. Although there is no noise in this numerical experiment, we have demonstrated in ref. 3 that, by using more physical constraints, ankylography can tolerate reasonably high noise as well as missing data at the center.

| Array size in Fourier space | $O_d$ | # of HIO iterations | *errorF* | *Error* |
|---|---|---|---|---|
| $101^3$ | 46.5 | 700 | 5.47e-4 | 2.2e-3 |
| $23^3$ | 2.18 | 700 | 3.29e-4 | 1.3e-3 |
| $19^3$ | 1.72 | 700 | 2.76e-4 | 1.5e-3 |
| $17^3$ | 1.14 | 2200 | 5.18e-5 | 3.13e-5 |

**Tab. 1** Parameters and the R-factors used for the ankylographic reconstructions of a 3D object.

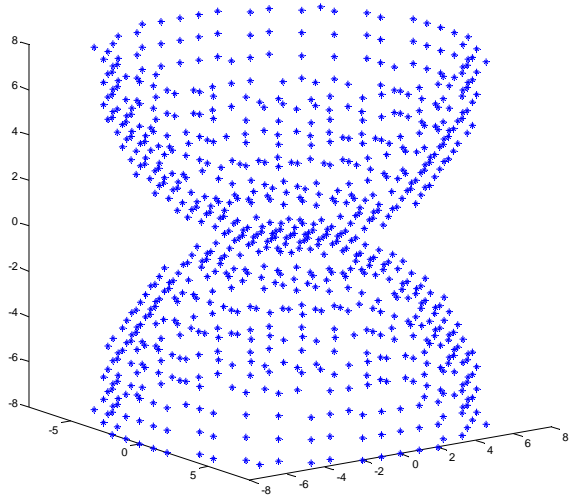

**Fig. 2** Distribution of the grid points within two hemi-spherical shells of 1 pixel thick. The array in Fourier space is $17^3$ voxels, the total number of non-centro-symmetrical grid points within a hemi-spherical shell is 393, and $O_d = 1.14$.

## II.     Explanation for the Experiment Reported in Ref. 3

In this section we further explain why we think that the Thibault's interpretation of our experiment is incorrect. Refs. 1 and 2 claimed that, due to the virtual opaqueness of our sample substrate (with tranmissivity of about $3.2 \times 10^{-4}$), there is dynamic scattering from our sample (*i.e.* the hollow mask region), which makes the Born approximation invalid. First of all, we agree that there is indeed a dynamic scattering effect in the direct beam that penetrates through the 100-nm-thick silicon nitride membrane. However, the magnitude of the directly penetrating wave is a few orders smaller than that of the scattered wave by the hollow mask region, and the directly penetrating beam is confined within a few pixels on the CCD detector. Thus compared to the diffraction pattern from the hollow mask, the direct wave penetrating through the membrane is negligible. Second, Thibault implied that the "physical origin" of the diffraction by the hollow mask is due to the edge scattering, which has a dynamic scattering effect [1,2]. We believe this statement is incorrect. If the far-field diffraction pattern is formed only by the waves scattered from the edges of the hollow mask, the reconstructed image from the diffraction pattern would only show the edges and there will be *no* density in the hollow region.

However, a number of groups have numerically and experimentally reconstructed the far-field diffraction patterns from apertures or masks (see, *e.g*., ref. 10). The reconstructed density is always continuous and covers the whole region of the apertures. Although there may be some density variation in the experimentally reconstructed images, it is mainly due to the incoherence of the incident beam and noise in the experiments.

Our interpretation of the experiment reported in ref. 3 is related to the quantum mechanics, as it is the coherent diffraction. In the famous Young's double-slit experiment, the interference pattern formed on the far-field screen is due to the fact that *one can't distinguish whether a photon is diffracted from slit 1 or 2* [11]. Likewise, in our coherent diffraction experiment, one just can't distinguish which point inside the hollow mask a photon comes from. Otherwise, there would be no interference pattern. Classically, this problem can be solved by using the Maxwell equations with boundary conditions [12]. Alternatively, one can use the classical Huygens-Fresnel theory as it can be directly derived from the Maxwell equations [13]. Based on the Huygens-Fresnel theory [14], we showed in appendix I that, when the edge effect is negligible, the far-field diffraction pattern from the 3D hollow mask is proportional to the square of the Fourier transform of the hollow mask sampled on the Ewald sphere.

Turn now to the multi-slice simulation results present in ref. 1. We believe they are inaccurate for the following reasons. First, the multi-slice formulation, first proposed by Cowley & Moodie in 1957 to deal with the multiple scattering effects in electron diffraction [15], is not rigorous. Here let us quote Cowley's descriptions of the multi-slice formulation in his classical book [16], "In the formulation of n-beam diffraction theory by Cowley and Moodie [1957] transmission of electrons through a sample is represented by transmission through a set of $N$ two-dimensional phase- and amplitude-objects separated by distance $\Delta z$. The total phase change and amplitude change of the electron wave in a slice of the specimen of thickness $\Delta z$ is considered to take place on one plane to the next is by Fresnel diffraction in vacuum. It has been shown (Modie [1972]) that in the limiting case that the thickness of the slice $\Delta z$ goes to zero and the number of slices $N$ goes to infinity in such a way that $N\Delta z = H$, where $H$ is the specimen thickness, this form of description becomes a rigorous representation of the scattering process, completely consistent with the more conventional quantum mechanical descriptions." Therefore, compared to the Huygens-Fresnel theory which remains the foundation to study the diffraction and scattering processes, the multi-slice formulation is not as accurate to analyze our experiment. Second, as clearly described in Cowley's above statement, the multi-slice formulation is a two-dimensional approximation, as the depth information is lost in the analysis. Thus it is incorrect to analyze a three-dimensional imaging technique (*i.e.* ankylography) by using a two-dimensional approximation method.

## III.    Appendix I

As per request by Thibault, we will use the Huygens-Fresnel theory to show that, when the edge effect is negligible, the far field diffraction intensities are proportional to the square of the Fourier transform of the 3D hollow mask sampled on the Ewald sphere. As it is the well-know derivation, we put it in the appendix. Fig. 3 illustrates two secondary waves generated from points O and P, where O is assumed to be the origin of the hollow mask. Based on the path difference, we obtain the phase shift of the two waves,

$$\varphi = -2\pi(\vec{S}_o - \vec{S}_i) \cdot \vec{r} = -2\pi\vec{k} \cdot \vec{r} \qquad (4)$$

where $\vec{S}_i$ and $\vec{S}_o$ are the incident and scattering wave vectors. By considering all the points inside the hollow mask and ignoring the edge effect, we obtain the structure factor in the far field

$$F(\vec{k}) = \int_V \rho(\vec{r})e^{-2\pi i\vec{k}\cdot\vec{r}} d^3\vec{r} \qquad (5)$$

where $\rho(\vec{r})$ is the 3D structure of the hollow mask. Based on the derivation in supplementary information (page 4 and supplementary Fig. 4) in ref. 3, we obtain the far-field diffraction intensities sampled on the Ewald sphere,

$$I(\theta,\varphi) \propto |F(\theta,\varphi)|^2 = \left| \int_V \rho(\vec{r})e^{-\frac{2\pi i}{\lambda}[x\sin\theta\cos\varphi + y\sin\theta\sin\varphi + z(\cos\theta-1)]} d^3\vec{r} \right|^2 \qquad (6)$$

which is Eq. (1) in ref. 3. According to Eq. (6), we conclude that the far field diffraction intensities of the hollow mask are proportional to the square of the Fourier transform of the 3D sample on the Ewald sphere. In other words, the Born approximation holds and the spherical diffraction pattern encodes the depth information.

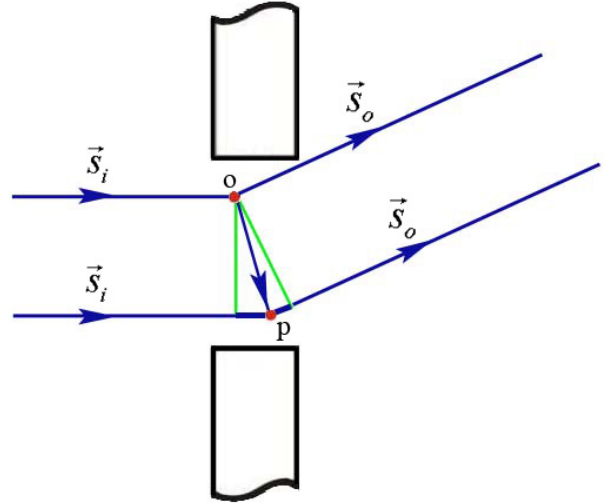

**Fig. 3** Secondary waves generated from points inside the 3D hollow mask. $\vec{S}_i$ and $\vec{S}_i$ are the incident and scattering wave vectors with $|\vec{S}_i| = |\vec{S}_o| = 1/\lambda$. Note that the dimensions are not to scale. In the experiment, the size of the 3D hollow mask is much larger than its thickness [3].

## Acknowledgements

We thank Huaidong for help with Fig. 3.